\documentclass[aps,pre,amsmath,amssymb,twocolumn,notitlepage,longbibliography,superscriptaddress]{revtex4-2}
\usepackage{graphicx}
\usepackage[colorlinks=true,allcolors=blue]{hyperref}
\usepackage{xcolor}
\usepackage{amsmath}
\usepackage{amsfonts}
\usepackage{amssymb}
\usepackage{varioref}
\usepackage{float}
\usepackage{dcolumn}   
\usepackage{subfigure}
\usepackage{bm}
\usepackage{multirow}
\usepackage[normalem]{ulem}
\usepackage{epstopdf}

\begin{document}

\title[]{Polymer rigidity controls conformational response to deformation-dependent nonreciprocal interactions}

	\author{Shalabh K. Anand}
	\affiliation{Institut f\"ur Theoretische Physik, Universit\"at Innsbruck, Technikerstrasse 21A, A-6020, Innsbruck, Austria}
	\email{shalabh.anand@uibk.ac.at}


\begin{abstract}
Nonreciprocal interactions, which violate action-reaction symmetry, constitute a broadly applicable mechanism for generating nonequilibrium forces in active and driven systems. Here, we explore how such interactions remodel polymer conformations using a minimal model in which local strain along the backbone gives rise to nonreciprocal forces. We find that bending rigidity is the decisive control parameter: flexible polymers contract under activity, whereas semiflexible polymers display a non-monotonic response that ultimately yields extension. These large-scale conformational changes arise despite only weak shifts in local structural statistics. Instead, activity reassigns bond-orientational correlations along the contour, damping orientational memory in flexible chains while amplifying it in semiflexible ones. Collectively, our results show that bending rigidity governs how deformation-dependent nonreciprocal activity transmits across length scales, offering a mechanism to regulate the structure and mechanics of active filamentous matter.
\end{abstract}

\maketitle

\section{Introduction}
The conformational and dynamical properties of polymers~\cite{howard2002mechanics,gupta2019morphological,ravichandran2017enhanced,smrek2017chromosom,saggiorato2015conformations} are governed by the interplay between chain connectivity, thermal fluctuations, and intramolecular interactions~\cite{fily2020buckling,chelakkot2012flow,chakrabarti2020flexible,das2021coil,biswas2021rigidity,sabeur2008kinetically,giomi2010statistical,kirkwood1948intrinsic}. In many biological and synthetic systems, however, polymers are driven out of equilibrium by internally generated forces, introducing additional couplings between polymer deformation and active stress generation.
Filamentous active matter~\cite{chelakkot2014flagellar,Anand2019,Anand2021,berg2003rotary,koenderink2009active,chelakkot2021synchronized,biswas2017linking,tiwari2020periodic,Nishiguchi2018} provides a fundamental framework for understanding how internally generated forces reshape the structure and dynamics of polymers far from equilibrium~\cite{winkler2020physics,winkler2017active,heeremans2022chromatographic,Mokhtari2019,Anand2024computer}. Biological filaments~\cite{heeremans2022chromatographic,miller2002emergence,zidovska2013chromosom,kemp2001helical,thampi2014instabilities}, including cytoskeletal networks~\cite{howard2002mechanics,harada1987sliding,ravichandran2017enhanced,ndlec1997self,kumar2018tunable} and chromosomes~\cite{butan2011spiral,berlatzky2008spatial,zidovska2013chromosom,saintillan2018chromosom,Goychuk2023,Beltran2019}, continuously consume chemical energy through molecular motors, while synthetic active materials exploit similar principles to generate autonomous motion and mechanical response. 

Equilibrium models treat polymers as elastically connected elements capturing thermal fluctuations and entropic elasticity. Yet many filamentous systems are intrinsically out of equilibrium~\cite{laskar2017filament,fily2020buckling,smrek2017chromosom,Goychuk2024}, driven by motors (Myosin II, Kinesin, Dynein) that convert chemical energy into forces sensitive to local extension, compression, or tension. At a coarse-grained level, such deformation-dependent force generation need not satisfy action-reaction symmetry, naturally giving rise to nonreciprocal interactions, producing persistent motion, large-scale reorganization, and anomalous mechanics.

Theoretical work on active polymers predominantly follows two routes~\cite{vandebroek2015dynamics,ghosh2014dynamics,abkenar2013collective,das2021coil}: (i) tangential activity along the filament, yielding polar active dynamics~\cite{isele2015self,Anand2018,philipps2022tangentially,tejedor2024progressive}; and (ii) active Brownian forces on monomers~\cite{kaiser2015does,Anand2020,eisenstecken2017internal,martin2018active,kaiser2014unusual}. These lead to distinct behaviors: polar models typically show a monotonic decrease of the radius of gyration with activity~\cite{Anand2018,fazelzadeh2023effects,van2024conformation}, whereas active Brownian polymers often display non-monotonic swelling-collapse transitions~\cite{Anand2020,eisenstecken2016conformational,Anand2025,anandselfalignment2026}. These studies have established how different forms of active forcing modify polymer conformations, but most existing models prescribe the active force independently of local polymer deformation. How a deformation-dependent violation of action-reaction symmetry couples to chain elasticity and bending rigidity remains largely unexplored.

Recent developments in nonequilibrium statistical physics have identified nonreciprocal interactions as a generic consequence of broken action-reaction symmetry~\cite{Morfill2009,Chiacchio2023,Soto2014,Ivlev2015,Bowick2022,Saha2020,Guislain2024,Kreienkamp2024,Gupta2022}, generating unusual collective motion~\cite{You2020,Duan2025,Ouazan-Reboul2023,Benois2023,Fruchart2021,Mandal2024,Banerjee2022}, directed transport and odd elasticity~\cite{Fruchart2023}, and anomalous mechanical responses~\cite{Hickey2023,Pisegna2024,Wittkowski2024}. Evidence points to their relevance in living, mechanically active environments with ongoing energy consumption and internal stress generation~\cite{Dadhichi2020,Pisegna2024,Blumenthal2024,Haluts2024,Yamanaka2014,Duan2023}.

Although nonreciprocal interactions have recently attracted considerable attention, their implications for polymer physics remain largely unexplored. Polymers provide a unique setting because local nonequilibrium forces can propagate through elastic connectivity over many contour lengths. Understanding this propagation is essential for identifying how local active processes reshape global conformations.

Here we introduce a minimal polymer model with strain-dependent nonreciprocal interactions. Each bond generates an active force proportional to its local deformation, coupling elasticity directly to nonequilibrium driving while explicitly breaking action-reaction symmetry. Using extensive numerical simulations, we show that bending rigidity fundamentally controls the conformational response to this activity. Flexible polymers compact whereas semiflexible polymers exhibit a non-monotonic extension. We identify the microscopic origin of this transition as a rigidity-dependent reorganization of orientational correlations that occurs despite only weak changes in local structural statistics. Our results establish deformation-dependent nonreciprocal activity as a distinct mechanism for controlling polymer conformations in active filamentous matter.

\section{Model}
\begin{figure}
    \centering
    \includegraphics[width=0.8\linewidth]{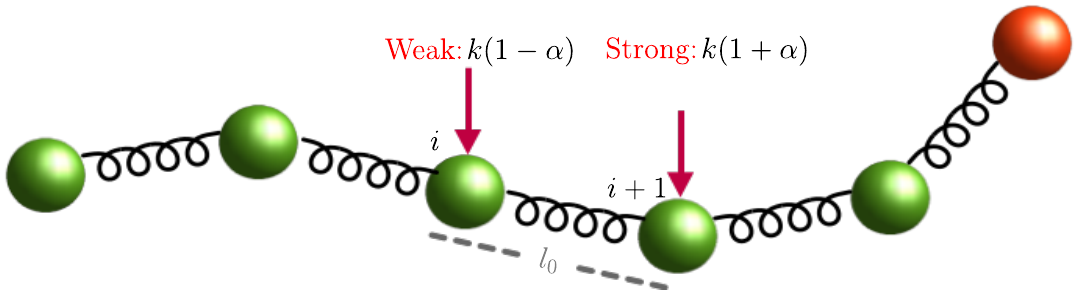}
    \caption{A schematic of the active polymer model. Here, the number of monomers are just a representation, the actual number of monomers are mentioned in the text. The front end of the polymer is represented in red monomer.}
    \label{fig:model}
\end{figure}

We model a three-dimensional linear polymer chain using a bead-spring model, where monomers are treated as beads connected via springs~\autoref{fig:model}. The spring potential for neighboring monomers is given by
\begin{equation}
    U_{s}(r) = \sum_{i=1}^{N-1} \frac{1}{2}k_{s}(\bm{R}_{i}-l_{0})^{2}.
    \label{Eq:spring_potential}
\end{equation}
Here, $k_{s}$ is the spring constant, $\bm{R}_{i} = \bm{r}_{i+1}-\bm{r}_{i}$ is the bond vector between the $i^{th}$ and $(i+1)^{th}$ monomers, and $l_{0}$ is the rest length of the spring. The positions of the monomers are represented by $\bm{r}_{i}$, where index $i$ labels the monomer.

Usually, the springs between successive monomers are considered reciprocal. However, we consider nonreciprocal springs to implement activity in this work~\cite{Kreienkamp2024,Hanai2024,Chiu2023}. The nonreciprocal interactions between $i$th and $j$th monomers results into a non-balacing pairwise force, i.e.,  $F_{sp}^{ij} \neq F_{sp}^{ji}$. In particular, for every pair of monomers $i$ and $j$ with $j=i+1$, the pairwise spring force could be written as,
\begin{eqnarray}
    \begin{aligned}
        F_{sp}^{ij} &= (1-\alpha) F_{c,sp}^{ij}, \\
        F_{sp}^{ji} &= (1+\alpha) F_{c,sp}^{ij}.
    \end{aligned}
    \label{eq:nonrec}
\end{eqnarray}
Here $\alpha$ is the nonreciprocal parameter, and $F_{c,sp}^{ij}$ is the conservative spring force computed from \autoref{Eq:spring_potential} and satisfies Newton's third law. Activity enters the model through nonreciprocal interactions that couple local bond deformation to force generation. For each bond connecting monomers $i$ and $i+1$, we assign different interaction strengths to its two ends, with a smaller force acting on monomer $i$ and a larger one acting on monomer $i+1$ as presented in \autoref{eq:nonrec}. Thus, each bond individually violates action-reaction symmetry while the underlying elastic connectivity remains unchanged. The same construction is applied to every bond along the polymer, with a fixed polarity directed toward the $N$th monomer. Consequently, an interior monomer receives nonreciprocal force contributions from both of its neighboring bonds, whereas the terminal monomers receive a contribution from only one bond.

We also implement excluded-volume interactions among the non-neighboring monomers of the chain via the Weeks-Chandler-Anderson (WCA) potential,
\begin{equation}
    U_{WCA}(r) = 4 \epsilon \Big[ \Big(\frac{\sigma}{r_{ij}} \Big)^{12} - \Big(\frac{\sigma}{r_{ij}} \Big)^{6} \Big] + \epsilon
\end{equation}
for inter-monomer distance $r_{ij}<2^{1/6}\sigma$, and $U_{WCA}=0$ for $r_{ij}>2^{1/6}\sigma$. Here, $\epsilon$ is the interaction strength and $\sigma$ is the diameter of the monomers. The bending energy is written as

\begin{equation}
    {U_b} = \frac{\kappa}{2} \sum_{i=1}^{N-3} ({\bm R}_{i+1} - {\bm R}_i)^{2}.
    \label{Eq:bending}
\end{equation}

The total energy thus consists of these three contributions, i.e., $U=U_s+U_{WCA}+U_b$. We use Brownian dynamics simulations to study the motion of the monomers of the polymer chain. The equation of motion for monomer $i$ is written as over-damped Langevin equation
\begin{equation}
    \gamma \dot{\bm{r}}_{i} = -\sum_{j\neq i}\nabla_{i}U_{ij} + \bm{f}^{r}_{i}(t).
\end{equation}
Here, $\gamma$ is the friction coefficient and $\bm{f}_i^r(t)$ is the thermal noise with zero mean and second moment
\begin{equation}
    \langle\bm{f}_i^r(t)\cdot\bm{f}_j^r(t')\rangle = 6k_BT\gamma\delta_{ij}\delta(t-t').
\end{equation}

All physical parameters are expressed in units of monomer diameter $\sigma$, the diffusion coefficient of a monomer $D_{m}$, and thermal energy $k_BT$. Unless mentioned otherwise, we fix $N=100$, $\epsilon=k_BT$, $\kappa$ in units of $k_BT/\sigma^2$, and keep $k_s=1000\,k_BT/\sigma^2$ and $l_0=\sigma$. The bending stiffness is defined such that $\kappa$ directly sets the persistence length.

\section{Results}

\begin{figure*}
    \centering
    \includegraphics[width=0.33\linewidth]{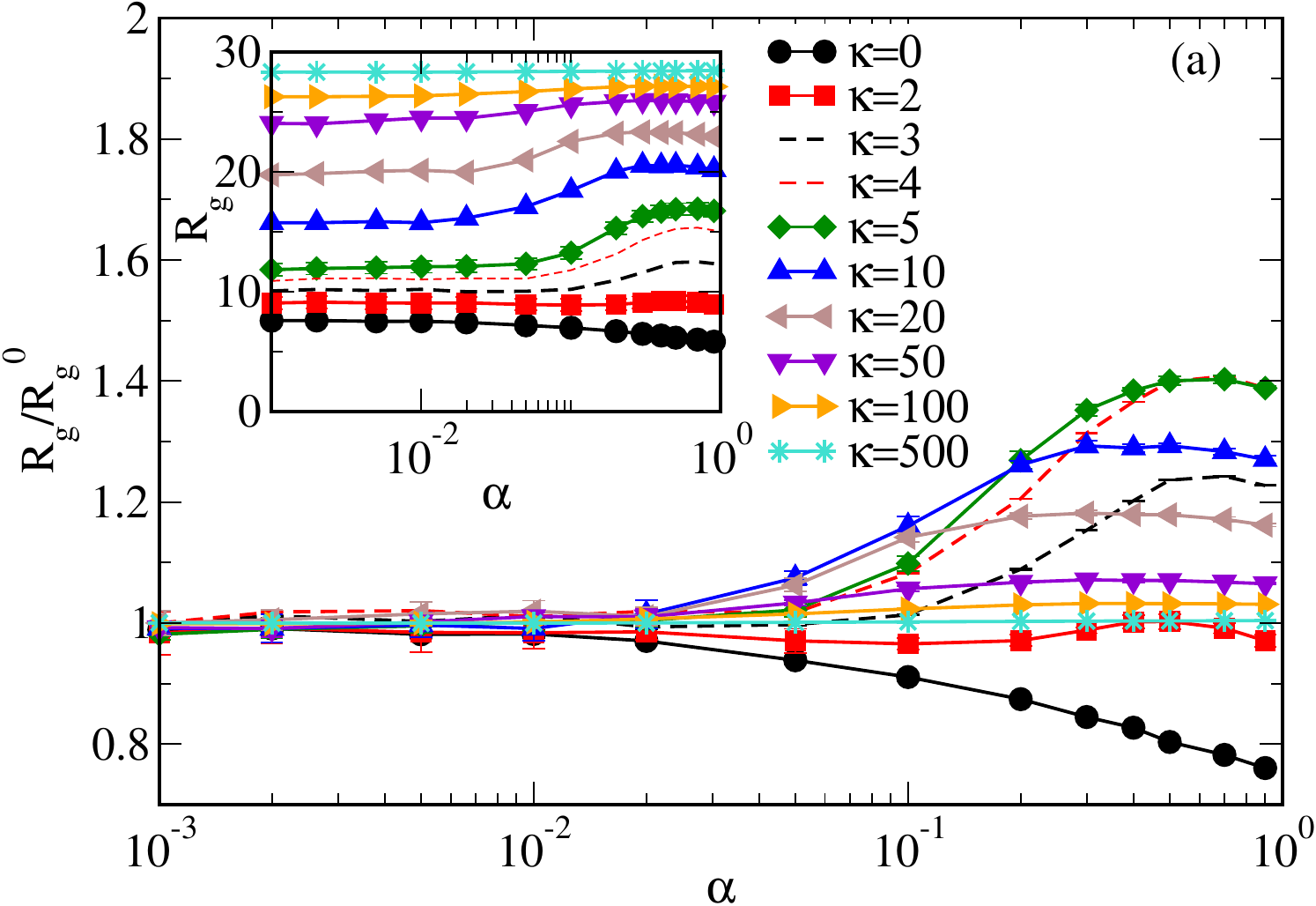}
    \includegraphics[width=0.315\linewidth]{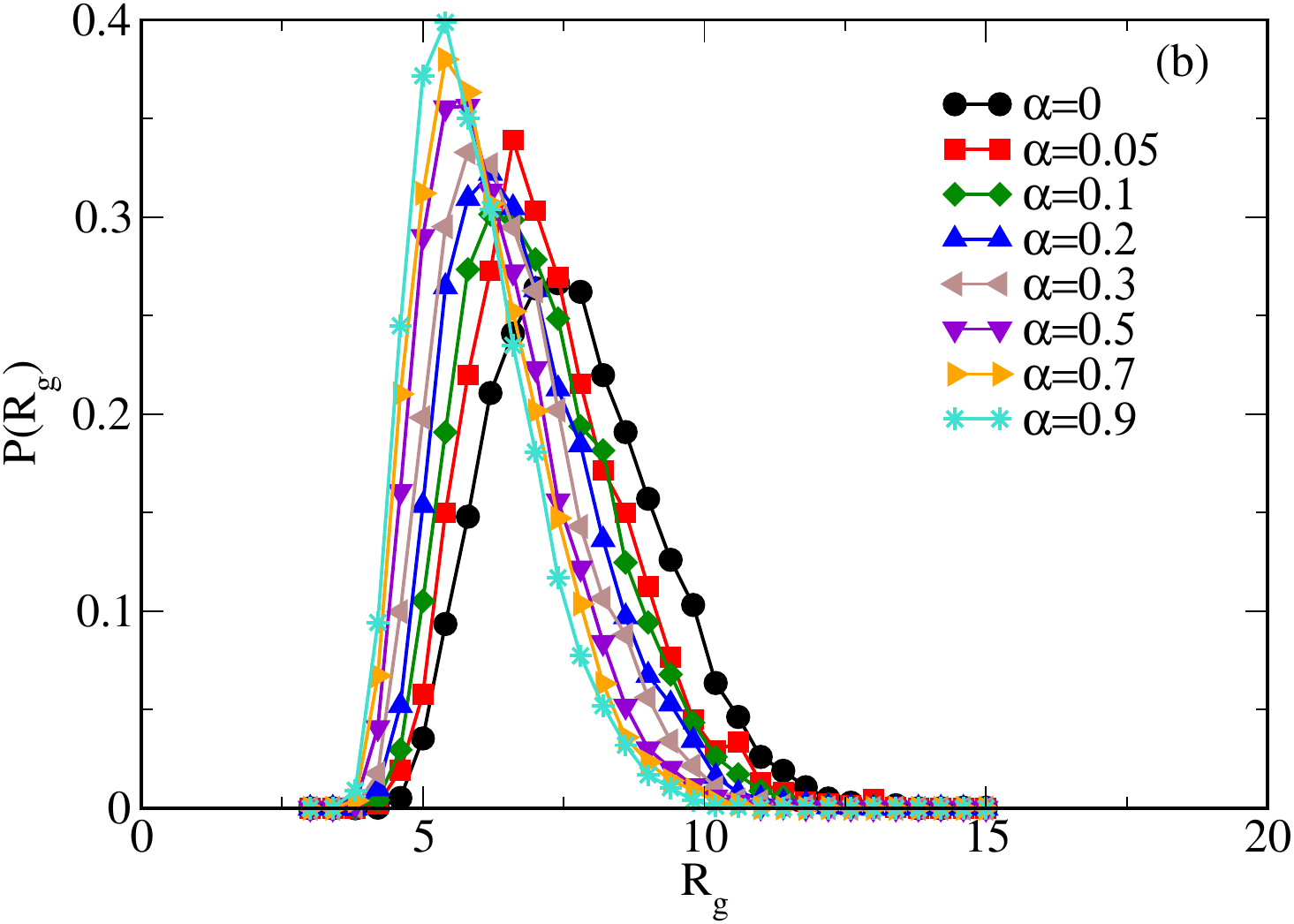}
    \includegraphics[width=0.315\linewidth]{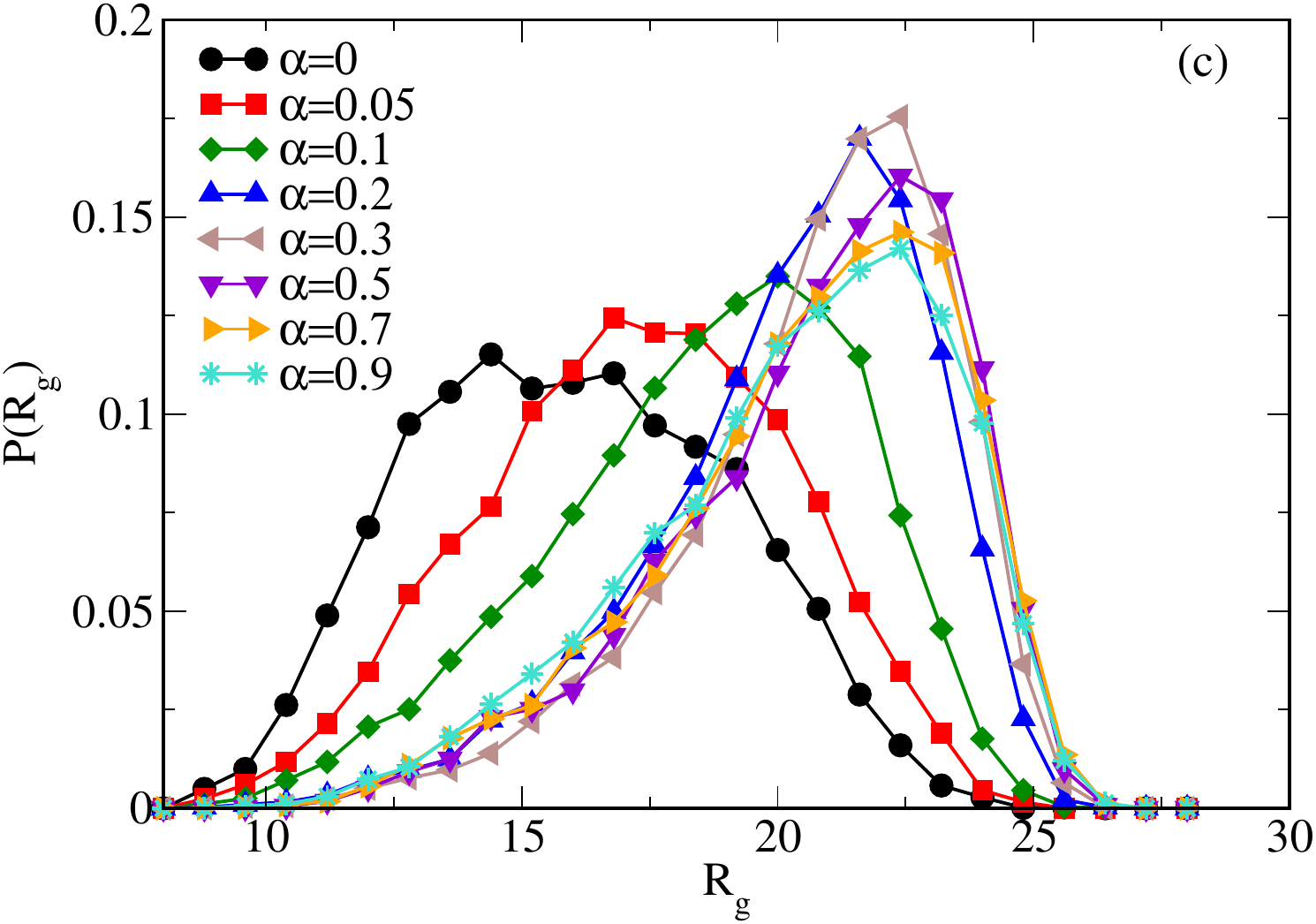}
    \caption{a) Average values of radius of gyration $R_g$ and distribution of $R_g$ for b) $\kappa=0$ and c) $10$ for different values of $\alpha$.}
    \label{fig:av-rg}
\end{figure*}

\subsection{Structural properties}
We now study the impact of nonreciprocal springs on the structural properties of the filament using the radius of gyration $R_g$ and end-to-end distance $R_e$, defined as
\begin{eqnarray}
    R_g &=& \sqrt{\Big\langle \frac{1}{N} \sum_{i=1}^{N} (\bm{r}_i - \bm{r}_{cm})^2 \Big\rangle}, \\
    R_e &=& \sqrt{\Big\langle (\bm{r}_N - \bm{r}_1)^2 \Big\rangle}.
\end{eqnarray}
Here, $\bm{r}_{cm}$ is the centre of mass of the filament, and the averaging is done over time and over different initial configurations. We present steady-state values of $R_g$ as a function of $\alpha$ for different bending stiffness parameters $\kappa$ in the inset of \autoref{fig:av-rg}a (see a similar pattern in $R_e$ in the supplementary text~\cite{Note1}). The $R_g$ increases with increasing $\kappa$ at any given $\alpha$, which is a consequence of the straightening of the filament due to the high bending rigidity. The radius of gyration of the polymer chain also varies with $\alpha$ for all rigidities, however, the variation is different in flexible and semiflexible regimes. To highlight the variation with $\alpha$ and $\kappa$, we scale the y-axis by the corresponding values at $\alpha=0$ in the main panel.

We first discuss the results for a flexible polymer chain, i.e., $\kappa=0$. At small activity strengths $\alpha$, there is a very little variation in $R_g$. However, $R_g$ decrease steadily upon further increasing $\alpha$. The radius of gyration decreases by almost $25\%$ at $\alpha=0.9$, indicating a very significant reduction in the structure. We now discuss the results for the semiflexible filament, i.e., $\kappa>0$. Even for a filament with bending stiffness $\kappa>0$, $R_g$ remain unchanged or weakly changed at low $\alpha$. This is the weak-activity regime, which is insufficient to drive structural changes in the filament. However, as $\alpha$ is further increased, the imbalance caused by nonreciprocal spring interactions plays an important role. The filament undergoes swelling, i.e., $R_g$ increases with $\alpha$. This trend changes its course of variation at high values of $\alpha$, where $R_g$ starts to decrease. Thus, a semiflexible filament displays non-monotonic structural behavior with $\alpha$. This is different from the active polar filament model~\cite{Anand2018,philipps2022tangentially,winkler2025conformational,Maurya2025} or an active Brownian filament model~\cite{kaiser2015does,Anand2020,kaiser2014unusual}. It is also worth noting that the filament undergoes more swelling for smaller $\kappa$ values: the increase in $R_e$ for $\kappa=5$ is larger than for $\kappa=10, 20, 50, 100$, and $500$. The degree of swelling is somewhat also dependent on $\kappa$. For clarity, we also show very small rigidities $\kappa=2$, $3$ and $4$, which resembles the behavior of a semiflexible filament. This is a prominent signature of the fact that even a very small bending rigidity changes the course of variation of the steady-state structural properties of the filament. 

We have unraveled a monotonic shrinkage of a flexible filament with $\alpha$ and a non-monotonic behavior (swelling followed by shrinkage) for a semiflexible filament. To characterize the changes in the structure of the filament in more detail, we plot the distribution of $R_g$ in \autoref{fig:av-rg}b and c. \autoref{fig:av-rg}b displays $P(R_g)$ for a flexible filament. The distribution shifts monotonically towards smaller $R_g$ values and becomes narrower with increasing $\alpha$, suggesting increasingly compact structures. For a semiflexible filament at $\kappa=10$ (see \autoref{fig:av-rg}c), the distribution shifts towards larger $R_g$ with $\alpha$. However, it also develops a broader tail towards small $R_g$ at sufficiently high $\alpha$, which drives the shrinkage in the average radius of gyration. Thus, it is not a completely compact structure that leads to shrinkage; rather, the filament dynamically explores both compact and extended configurations. For a high value of $\kappa$ (say $100$), the distribution shifts towards larger $R_g$ and then stalls as $\alpha$ increases (see the supplement~\cite{Note1}), consistent with swelling followed by near-saturation of the average structural properties.

\begin{figure}
    \centering
    \includegraphics[width=0.9\linewidth]{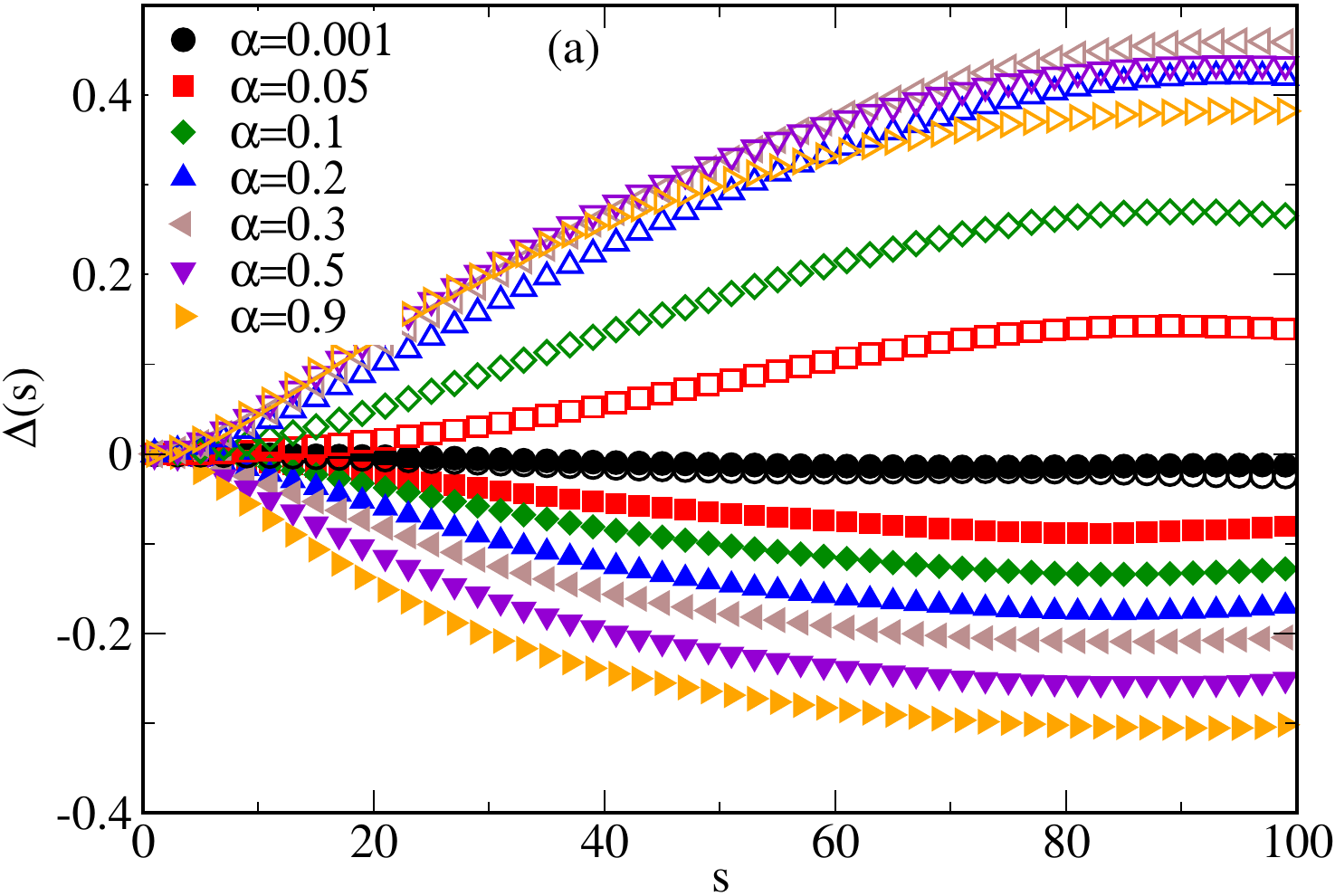}
    \includegraphics[width=0.9\linewidth]{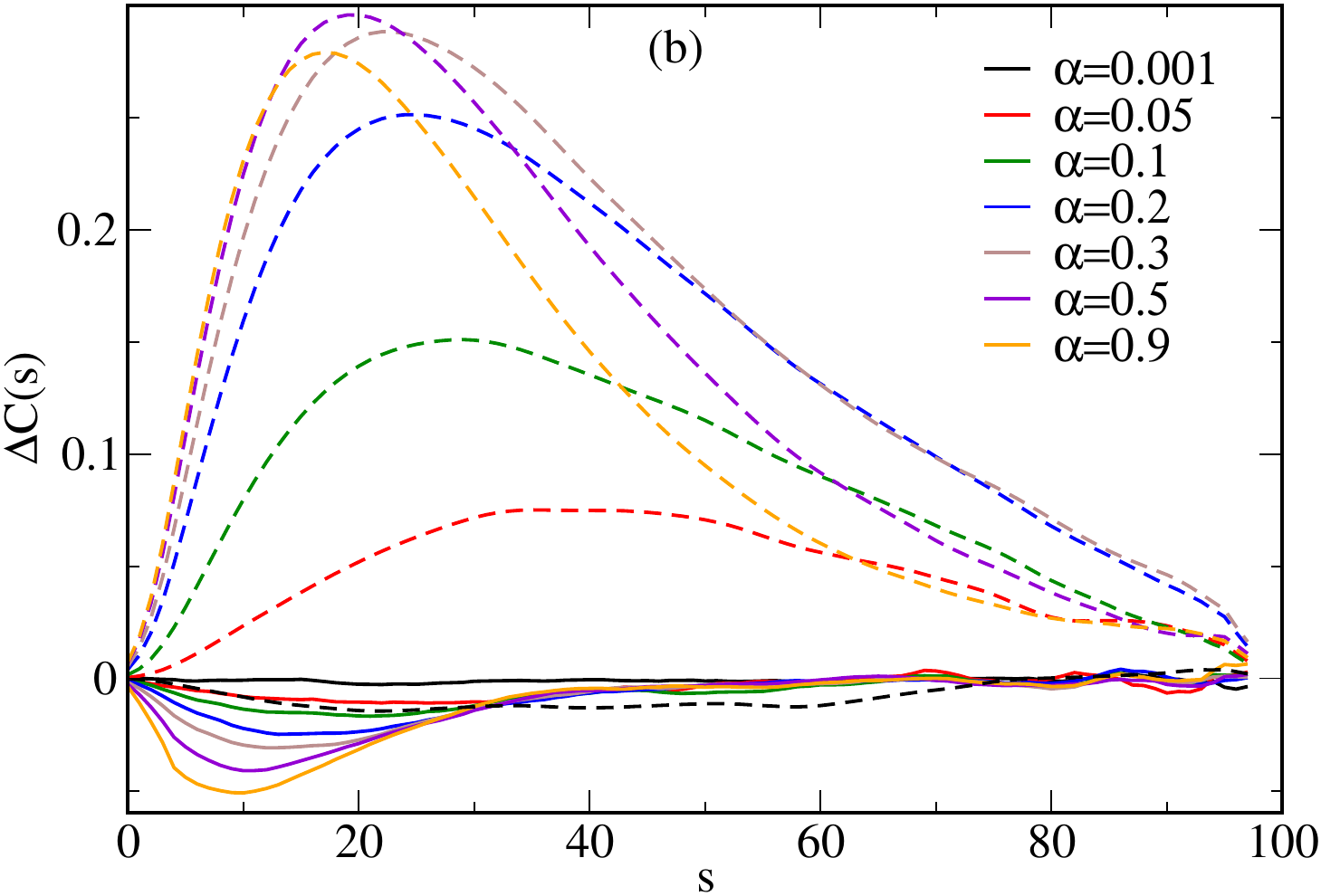}
    \caption{a) Response function $\Delta(s)$ as a function of arc-distance $s$. Here filled symbols represent the data for $\kappa=0$, and open symbols for $\kappa=10$. b) Difference between the bond-correlation with respect to its passive counterpart $\Delta C(s)$ for $\kappa=0$ (solid lines) and $\kappa=10$ (dotted lines).}
    \label{fig:mechanism}
\end{figure}

The opposite response of flexible and semiflexible polymers suggests that nonreciprocal activity modifies the internal organization of the polymer in a manner that cannot be understood from local structural measures alone. Indeed, the bond-length distribution, spring energy, and bending energy exhibit only weak changes with activity (Supplementary Information), indicating that the origin of the transition lies in the long-range organization of the polymer contour. We therefore show inter-monomer distance $R(s) = |\mathbf{r}_{i+s}-\mathbf{r}_{i}|$ (see supplement in \cite{Note1}), and present a kind of response function $\Delta (s)$ in \autoref{fig:mechanism}a, which gives us information about how monomers arrange themselves with respect to their passive counterpart,
\begin{equation}
    \Delta(s) = \frac{R(s,\alpha)-R(s,0)}{R(s,0)}.
    \label{eqn:response}
\end{equation}
The variation of $\Delta (s)$ for flexible polymers exhibit a negative response over long contour distances (filled symbols for $\kappa=0$ in \autoref{fig:mechanism}a), whereas semiflexible polymers develop a positive and non-monotonic response (open symbols for $\kappa=10$ in \autoref{fig:mechanism}a).

This raises a very fundamental question: What determines this opposite behavior? Since the end-to-end distance is exactly determined by bond correlations~\cite{rubinstein2003polymer},
\begin{equation}
    \langle R_e^2 \rangle = l_0^2 \Big[ N + 2\sum_{s=1}^{N-1} (N-s)C_b(s) \Big].
\end{equation}
The microscopic origin of the transition must be encoded in the activity-induced modification of $C_b(s)$ (see Supplementary Information). We rather look at the difference between the bond-correlation and its passive counterpart $\Delta C(s)$ in \autoref{fig:mechanism}b,
\begin{equation}
    \Delta C(s) = C_b(\alpha,s) - C_b(0,s).
\end{equation}
The activity-induced change in the bond correlation is qualitatively different in the two regimes. For flexible polymers, $\Delta C(s)$ is negative over the entire contour (solid lined for $\kappa=0$), indicating a suppression of orientational correlations. In contrast, $\Delta C(s)$ remains positive for semiflexible polymers (dotted lines for $\kappa=10$), demonstrating that bending rigidity converts the same nonreciprocal driving into enhanced orientational coherence.

\begin{figure}
    \centering
    \includegraphics[width=0.9\linewidth]{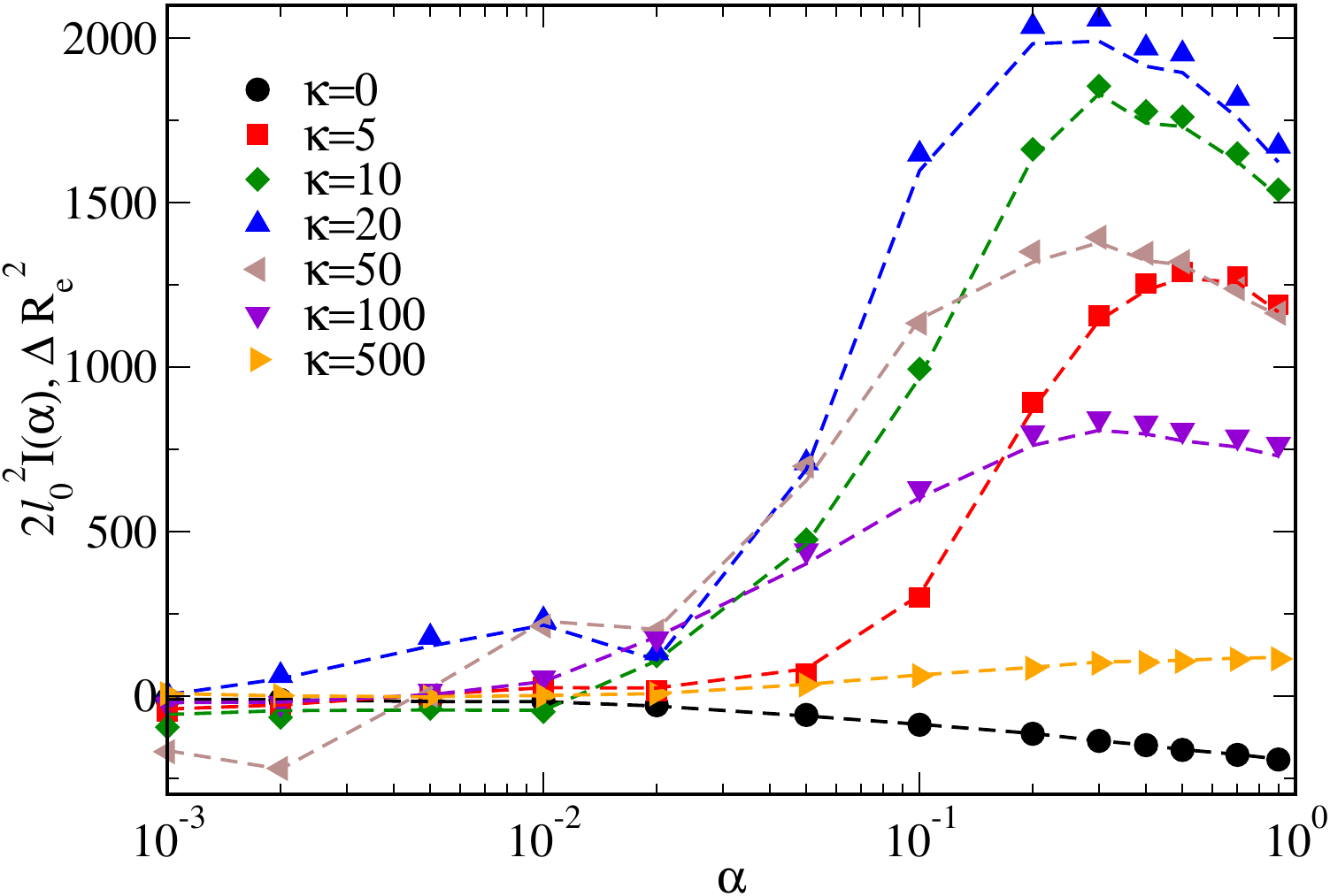}
    \caption{Cumulative weighted bond-correlation $I(\alpha)$ as a function of $\alpha$ for different values of bending rigidity $\kappa$.}
    \label{fig:predict}
\end{figure}
To quantify how the redistribution of orientational correlations influences the global conformation, we define
\begin{equation}
    I(\alpha) = \sum_{s=1}^{N-1} (N-s) \Delta C(s),
\end{equation}
whose weighting follows directly from the exact expression for $\langle R_e^2\rangle$. We show $I(\alpha)$ as a function of $\alpha$ for different values of $\kappa$, starting from flexible to rigid regime in \autoref{fig:predict}, which captures the variation of structural properties for both flexible and semiflexible polymers. We present the $\Delta R_e^2 = (\langle R_e^2 (\alpha) \rangle - \langle R_e^2 (0) \rangle)$ in \autoref{fig:predict} in dotted lines, which matches very well with $I(\alpha$). These results reveal that nonreciprocal activity acts primarily by reorganizing orientational correlations rather than local structure. Flexible polymers respond by suppressing orientational memory, leading to chain compaction. In contrast, bending rigidity converts the same active driving into enhanced orientational correlations, producing chain extension. Thus, bending rigidity fundamentally changes how nonreciprocal activity propagates along the polymer contour and determines the macroscopic conformational response.

\begin{figure}
    \centering
    \includegraphics[width=0.9\linewidth]{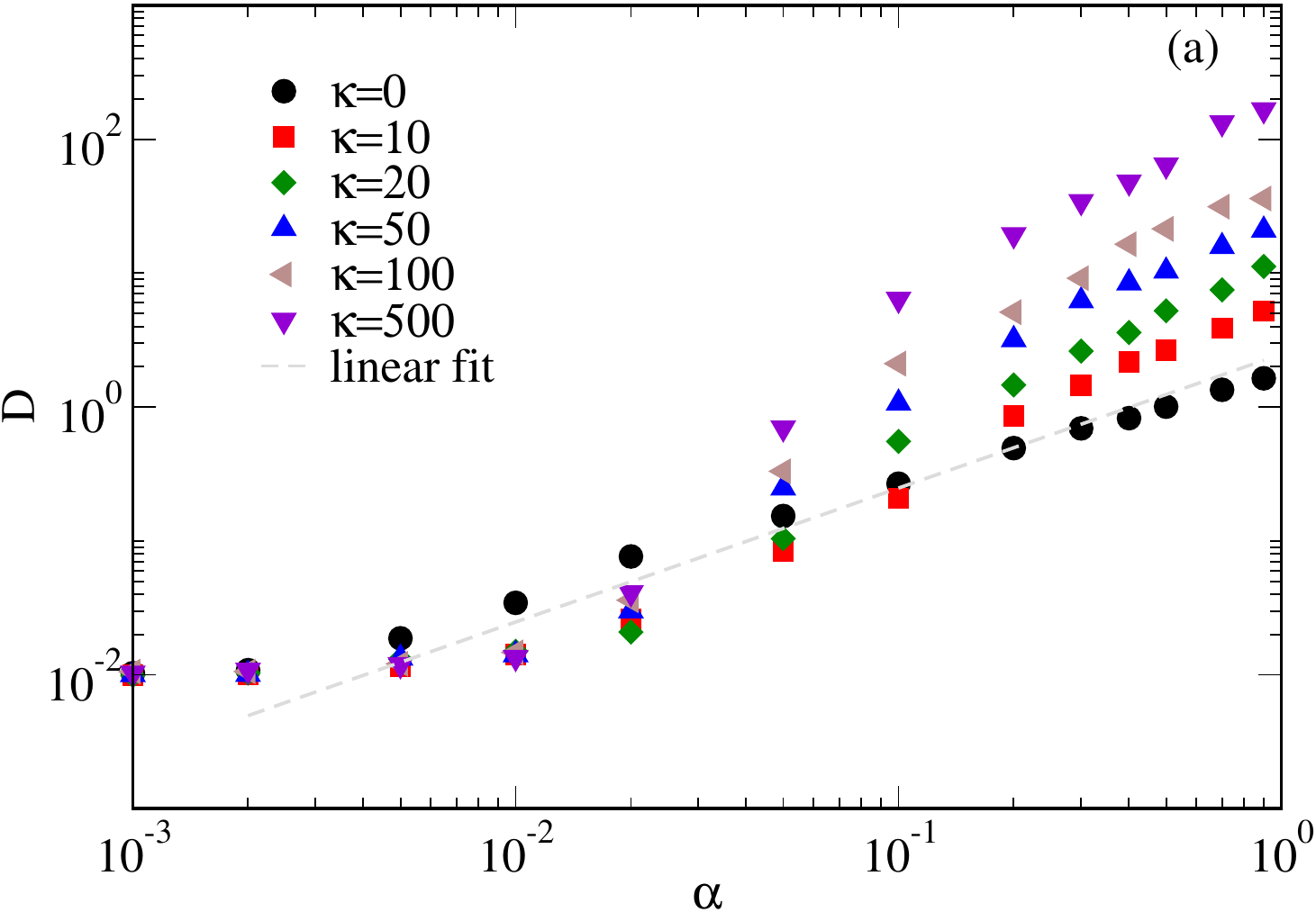}
    \includegraphics[width=0.9\linewidth]{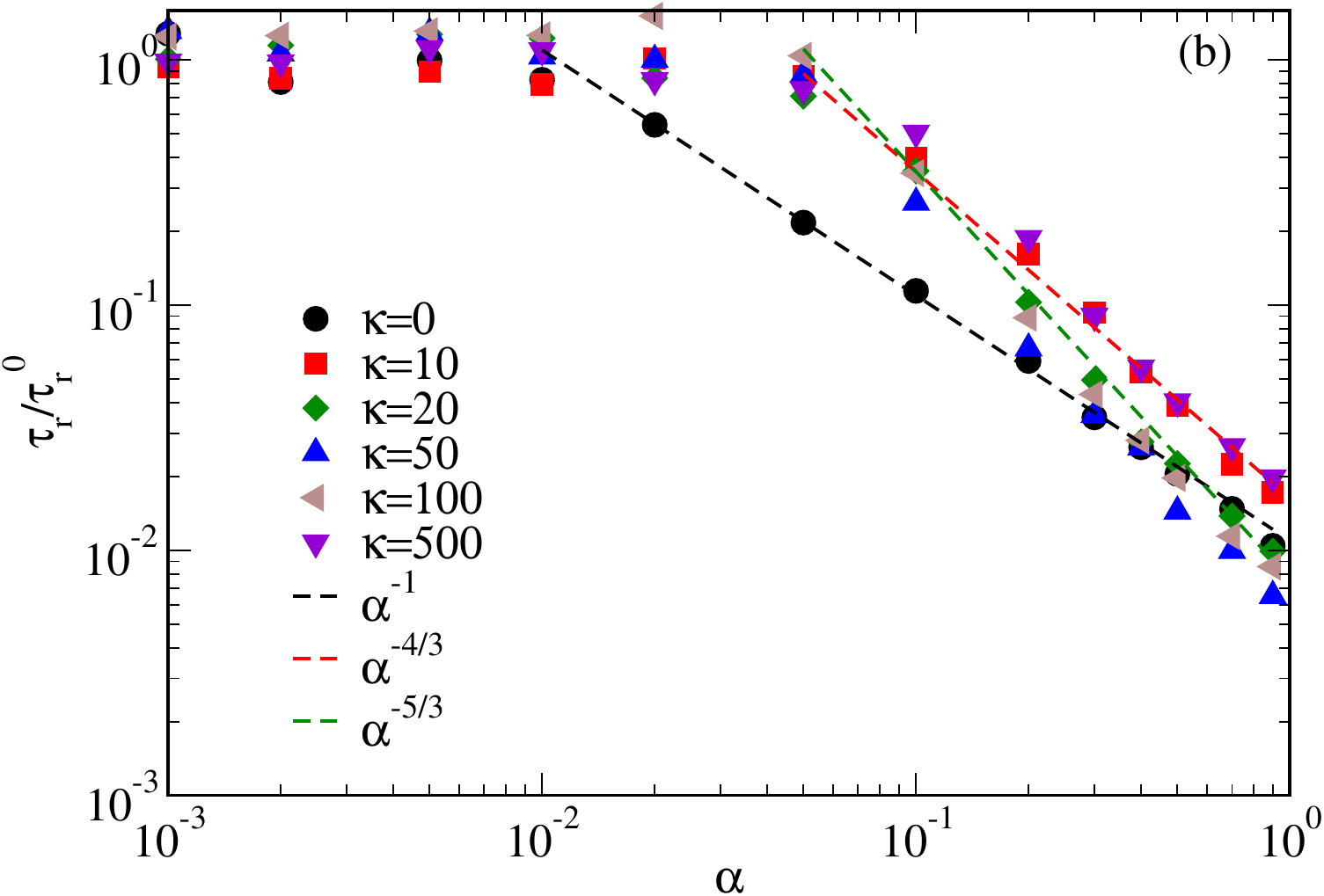}
    \caption{Dynamics of the filament: a) Long-time diffusion constant computed from the linear regime of the MSD of the centre of mass, as a function of $\alpha$ for different bending stiffnesses. b) Relaxation time computed from the end-to-end vector correlation ($C_e(t) \sim \exp(-t/\tau_r)$), scaled by the value at $Pe=0$.}
    \label{fig:dynamics}
\end{figure}
\subsection{Dynamics}
Having established that nonreciprocal activity primarily reorganizes orientational correlations along the polymer contour while leaving the local structure nearly unchanged, we now examine how this microscopic reorganization influences the filament dynamics. We characterize the translational motion through the mean-squared displacement (MSD) of the center of mass, $\Delta_{CM}^{2}(t) = \langle(\bm{r}_{CM}(t)-\bm{r}_{CM}(0))^2\rangle$. The MSD of a polymer chain exhibits ballistic behavior in the short-time regime, i.e., $\Delta_{CM}^2(t) \sim t^2$, followed by diffusive behavior in the long-time regime, $\Delta_{CM}^2(t)=6Dt$, where $D$ is the long-time diffusion coefficient.

The variation of $D$ with the nonreciprocal strength $\alpha$ is shown in \autoref{fig:dynamics}a. The diffusivity increases monotonically with both activity and bending rigidity. While the flexible polymer displays an approximately linear increase with $\alpha$, progressively stiffer filaments exhibit a super-linear dependence. This behavior reflects the increasingly coherent propagation of nonreciprocal activity along the filament as bending rigidity enhances orientational correlations.

To probe the internal dynamics, we compute the end-to-end vector autocorrelation, $C_e(t) = \langle\bm{R}_e(0)\cdot\bm{R}_e(t)\rangle \simeq \exp(-t/\tau_r)$. Here, $\tau_r$ is the longest relaxation time. \autoref{fig:dynamics}b shows the normalized relaxation time $\tau_r(\alpha)/\tau_r(0)$ as a function of $\alpha$. The relaxation time decreases monotonically with activity and follows a power-law decay at sufficiently large $\alpha$. The decay exponent depends strongly on bending rigidity: the flexible chain exhibits $\tau_r \sim \alpha^{-1}$, consistent with the active polar filament model, whereas increasingly stiff filaments relax substantially faster under nonreciprocal driving. Together with the enhanced diffusivity, these results demonstrate that the activity-induced reorganization of orientational correlations modifies not only the equilibrium conformation of the filament but also its long-time transport and relaxation dynamics.

\section{Summary and Discussion}
We develop a minimal model for active polymers that preserves elastic connectivity yet converts local deformation into strain-dependent, nonreciprocal forces. Departing from conventional approaches, the activity couples directly to the filament’s local deformation, establishing a minimal framework for deformation-dependent active stress in driven filamentous systems.

Our computational analysis shows that nonreciprocal activity generates qualitatively distinct conformational responses conditioned on polymer rigidity: flexible chains compact under activity, whereas semiflexible chains show a non-monotonic trajectory that ends in chain extension. These macroscopic changes arise despite only weak alterations in local structural quantities, including the bond-length distribution, spring energy, bending energy, and curvature statistics. Collectively, this indicates that nonreciprocal activity mostly reshapes long-range conformational correlations instead of significantly perturbing local filament structure.

By analyzing the bond-orientational correlation function, we uncover the underlying mechanism. Nonreciprocal activity diminishes orientational correlations in flexible polymers, but reverses course and strengthens them when bending rigidity is introduced. Consequently, bending rigidity qualitatively reshapes the way nonreciprocal activity propagates orientational information along the chain, leading to opposite conformational responses in flexible and semiflexible polymers. The resulting variations in overall polymer size align with the activity-driven reweighting of contour correlations implied by the exact identity linking bond correlations to the end-to-end distance.

Our findings pinpoint a simple microscopic pathway by which deformation-dependent, nonreciprocal activity propagates through filamentous materials. Rather than inducing substantial local structural rearrangements, this activity reweights and redistributes orientational correlations along the polymer contour, thereby allowing bending rigidity to determine the emergent global conformation. In turn, rigidity is established as the principal control parameter governing the structural response of polymers to nonreciprocal activity.

The proposed framework naturally lends itself to several future directions. On the theoretical side, it will be important to develop continuum descriptions that relate deformation-dependent nonreciprocal interactions to effective active elasticity and transport coefficients. It also provides experimentally testable predictions for synthetic active polymers and motor-driven filament systems in which active stresses depend on local deformation. It will also be interesting to examine how the proposed mechanism influences collective behavior in polymer networks and cross-linked active gels. More broadly, our findings demonstrate that bending rigidity fundamentally controls how deformation-dependent nonreciprocal activity propagates across length scales, providing a general route for programming the structure and dynamics of active filamentous matter.

\section{Acknowledgments}
We thank Jaime Agudo-Canalejo for critically reading the manuscript and providing his feedback. HPC facility LEO is acknowledged for providing computational resources.

\bibliography{references}

\end{document}